\documentclass[aps,prc,superscriptaddress,reprint]{revtex4-1}
\usepackage{graphicx} 
\usepackage[latin1]{inputenc}
\UseRawInputEncoding 
\usepackage{epstopdf}
\usepackage{dcolumn} 
\usepackage{bm}
\usepackage{hyperref}
\usepackage{array}
\usepackage{setspace}
\usepackage{booktabs}
\usepackage{amsmath}
\usepackage{mathtools}
\usepackage[section]{placeins}
\hypersetup{
    colorlinks=true,
    linkcolor=blue,
    filecolor=gray,
    urlcolor=blue,
    citecolor=blue,
}
\begin{document}
\title{Production of neutron-rich actinide nuclides in isobaric collisions via multinucleon transfer reactions}
\author{Peng-Hui Chen}
\email{Corresponding author: chenpenghui@yzu.edu.cn}
\affiliation{School of Physical Science and Technology, Yangzhou University, Yangzhou 225009, China}
\affiliation{Institute of Modern Physics, Chinese Academy of Sciences, Lanzhou 730000, China}

\author{Chang Geng}
\affiliation{School of Physical Science and Technology, Yangzhou University, Yangzhou 225009, China}
\author{Hao Wu}
\affiliation{School of Physical Science and Technology, Yangzhou University, Yangzhou 225009, China}


\author{Xiang-Hua Zeng}
\affiliation{School of Physical Science and Technology, Yangzhou University, Yangzhou 225009, China}
\affiliation{College of Electrical, Power and Energy Engineering, Yangzhou University, Yangzhou 225009, China }

\author{Zhao-Qing Feng}
\email{Corresponding author: fengzhq@scut.edu.cn}
\affiliation{School of Physics and Optoelectronics, South China University of Technology, Guangzhou 510641, China}

\date{\today}
\begin{abstract}
We have calculated the multinucleon transfer reactions of $^{208}$Os, $^{208}$Pt, $^{208}$Hg, $^{208}$Pb,$^{208}$Po, $^{208}$Rn, $^{208}$Ra,$^{132,136}$Xe bombarding on $^{232}$Th and $^{248}$Cm at Coulomb barrier energies within the dinuclear system model, systematically. The results are in good agreement with the available experimental data.
Coulomb effect and shell effect on production of actinides in these reactions have been investigated thoroughly.
Potential energy surface and total kinetic energy mass distributions in the reactions $^{208}$Hg, $^{208}$Pb and$^{208}$Po colliding on $^{248}$Cm and $^{232}$Th are calculated and analyzed, respectively. It is found that PES and TKE spectra manifest the fragment formation mechanism in the multinucleon transfer reactions. The isospin effect and shell effect are shown in PES and TKE. 
Production cross-sections of multinucleon transfer products are highly dependent on the isobar projectiles with mass number $A=208$.
The isobar projectiles with larger N/Z ratios are favorable for creating the neutron-rich target-like fragments. The isobar projectiles with larger charge number induced products tend to shift to proton-rich region. Coulomb potential coupled to shell effect is shown in production cross-sections of actinide isotopes.
Based on the radioactive projectiles induced reactions, we have predicted massive new actinide isotopes around nuclear drip lines, even could access the superheavy nuclei region.

\begin{description}
\item[PACS number(s)]
25.70.Jj, 24.10.-i, 25.60.Pj
\end{description}
\end{abstract}

\maketitle

\section{Introduction}
So far, including the synthesis of $^{149}$Lu \cite{PhysRevLett.128.112501}, $^{150}$Yb \cite{PhysRevLett.127.112501,nodate} and $^{207}$Th\cite{PhysRevC.105.L051302} in this year, there are 3327 species nuclei existed in the nuclide chart as we known, which consist of 288 natural nuclides (254 stable isotopes longer-lived than the earth and 34 unstable nuclides) and 3039 species nuclei synthesized in laboratories over the world based on methods of fusion-evaporation (FE), multinucleon transfer (MT) or deep inelastic reactions (DIR), projectile fragmentation (PF), spallation, fission (SF), neutron capture (NC), thermonuclear test (TT)\cite{Thoennessen2016}. However, there may be 8000-1000 unknown bounded isotopes predicted to exist by some theoretical models\cite{ZHANG2022101488,MOLLER20161,JACHIMOWICZ2021101393} in the whole nuclear map. Therefore, at least, over 5000 nuclides are waiting to be produced in laboratories by theorist and experimentalist, especially in the regions of nuclear drip lines and stability island of superheavy nuclei.

In recent years, from the experiment side, laboratories all over the world have synthesiszed saveral new species nuclei such as $^{207}$Th, $^{235}$Cm, $^{214}$U, $^{222}$Np, $^{211}$Pa, $^{280}$Ds\cite{PhysRevLett.126.032503,PhysRevC.105.L051302,PhysRevLett.126.152502,PhysRevC.102.044312} $etc.$ produced by FE reactions, $^{110}$Zr, $^{121}$Tc and $^{129}$Pd $etc.$ produced by PF\cite{doi:10.7566/JPSJ.87.014202}, $^{223,229}$Am and $^{233}$Bk \cite{DEVARAJA2015199} $etc.$ produced by MNT. 
It draws lots of interests, for Lanzhou Heavy Ion Research Facility (HIRFL) in China, Joint Institute for Nuclear Research (JINR) in Russia, Helmholtz Centre for Heavy Ion Research (GSI) in Germany and Grand Acc{\"e}l{\"e}rateur National d'Ions Lourds (GANIL) in France and Argonne national laboratory (ANL) in America, to synthesize new nuclides around drip lines and superheavy region. 

In order to describe the damped collision mechanism and predict synthesis cross-sections of the objective nuclides, theorists have built some sophisticated and practical models to depict the multinucleon transfer reactions at incident enerfy near the Coulomb barrier. For example, the GRAZING model\cite{HESSBERGER1994121,WINTHER1995203}, the dinuclear system (DNS) model\cite{PhysRevC.80.067601,WINTHER1994191}, and a dynamical model based on the Langevin equations\cite{ZAGREBAEV2015257,2018How}. Microscopic methods based on the degree of freedom of nucleons include the Time-dependent Hartree-Fock (TDHF) approach\cite{PhysRevLett.103.042701,PhysRevLett.106.112502,Jiang_2018}, and the improved quantum molecular dynamics model (ImQMD)\cite{PhysRevC.77.064603,PhysRevC.88.044605}.
Generally, all of these models could nicely reproduce the available experimental data through their unique characteristics. 
The dinuclear system model (DNS) can better consider shell effect, dynamical deformation, fission, quasi-fission, deep-inelastic mechanisms and odd-even effect, and its calculation efficiency is very high. 

In this work, the reactions of $^{132,136}$Xe + $^{248}$Cm at incident energy around Coulomb barriers have been compared to the available experimental data based on the DNS model, initially. 
Our calculations and experimental results have a nicely consistent trend.
We are focus on production cross-sections of exotic actinides in isobaric projectiles around Z = 82 induced MNT reactions with targets $^{232}$Th and $^{248}$Cm near Coulomb barrier. Based on calculation cross-sections of MNT fragmnts, Coulomb coupled to shell effect has been investigated thoroughly.
The article is organized as follows: In Sec. \ref{sec2} we give a brief description of the DNS model. Calculated results and discussions are presented in Sec. \ref{sec3}. Summary is concluded in Sec. \ref{sec4}.

\section{Model description}\label{sec2}

Initially, the DNS concept was proposed by Volkov for depicting the deep inelastic heavy-ion collisions\cite{VOLKOV197893}. G.G. Adamian added quasifission component in massive fusion process\cite{ADAMIAN1997361,ADAMIAN1998409}. Finally, the modifications of the relative motion energy and angular momentum of two colliding nuclei coupling to nucleon transfer within the DNS concept were developped by the Lanzhou Group \cite{PhysRevC.76.044606}. 
The cross sections of SHN, quasi-fission and fusion-fission dynamics have been extensively investigated within the dynamical DNS model. The dynamical evolution of colliding system sequentially proceeds the capture process by overcoming the Coulomb barrier to form the DNS, relaxation process of the relative motion energy, angular momentum, mass and charge asymmetry $etc$. within the potential energy surface and the de-excitation of primary fragments\cite{FENG200650}.
 The production cross section of the MNT fragments were evaluated by
\begin{eqnarray}
&&\sigma_{\rm tr}(Z_{1},N_{1},E_{c.m.})=\sum_{J=0}^{J_{\rm max}} \sigma_{cap}(E_{\rm c.m.},J)\int f(B)  \nonumber \\
&&\times P(Z_{1},N_{1},J_{1},B)  
\times W_{\rm sur}(E_{1},J_{1},s) dB.
\end{eqnarray}
The $\sigma_{\rm cap}(E_{\rm c.m.},J)$ is the cross-sections of DNS formation derived by Hill-Wheeler formula with barrier distribution function\cite{PhysRevC.101.024610}. 
$W_{\rm sur}(E_1,J_1,s)$ is the survival probability of fragments formation in MNT process. The s stands for the decay channels for fragments $(Z_{1},N_{1})$, such as neutron, proton, deuteron, alpha, gamma rays etc.
$E_{\rm c.m.}$ is incident energy in centre of mass frame.
The largest angular momentum $J_{\rm max}$ were calculated at the grazing configuration for colliding system.
The angular momentum $J$ is taken at initial coliding configuration before dissipating. 
$E_{1}$ and $ J_{1}$ represent the excitation energy and angular momentum for the fragment with proton number $Z_{1}$ and neutron number $ N_{1}$ in dinuclear system model (DNS), respectively. 
$P(Z_{1},N_{1},J_{1},B)$ is formation probability of fragments $(Z_{1},N_{1})$.
For the barrier distribution function, we take the asymmetry Gaussian\cite{Chen_2016} form.
\begin{eqnarray}
f(B)= \frac{1}{N}\exp\left [ - \right (\frac{B-B_{m} }{\bigtriangleup} )^{2}] 
\end{eqnarray}

The quantities $\bigtriangleup$ and $B_{\rm m}$ were evaluated by $\bigtriangleup$ =($B_{\rm t}$+$B_{\rm s}$)/2, $B_{\rm m}$=($B_{\rm t}$+$B_{\rm s}$)/2. $B_{\rm t}$ and $B_{\rm s}$ represent the height of coulomb barrier and the minimum point of deformation under tip-tip collision respectively. The normalization constant satisfies $\int f(B)dB=1$

In DNS model, the solution of nucleon transfer and relative motion carries out a set of microscopic derivations, master equations distinguish proton and neutron. The fragments distribution probability, P($Z_{1},N_{1},E_{1}$) represents the proton number $Z_{1}$, neutron number $N_{1}$, and excitation energy $E_{1}$ for DNS fragment 1 is described by the following master equation

\begin{eqnarray}
\label{mst}
&&\frac{d P(Z_1,N_1,E_1,\beta,t)}{d t} =  \nonumber \\ 
&&  \sum \limits_{Z^{'}_1}  W_{Z_1,N_1,\beta;Z'_1,N_1,\beta}(t) [d_{Z_1,N_1} P(Z'_1,N_1,E'_1,\beta,t) \nonumber   \\
&& - d_{Z'_1,N_1}P(Z_1,N_1,E_1,\beta,t)]   \nonumber   \\
&& + \sum \limits_{N'_1}  W_{Z_1,N_1,\beta;Z_1,N'_1,\beta}(t) [d_{Z_1,N_1}P(Z_1,N'_1,E'_1,\beta,t) \nonumber   \\ 
&& - d_{Z_1,N'_1}P(Z_1,N_1,E_1,\beta,t)]
\end{eqnarray} 
The $W_{Z_{1},N_{1},\beta;Z^{'}_{1},N_{1},\beta}$($W_{Z_{1},N_{1},\beta,;Z_{1},N^{'}_{1},\beta}$) is the mean transition probability from the channel($Z_{1},N_{1},E_{1},\beta$) to ($Z^{'}_{1},N_{1},E^{'}_{1},\beta$), [or ($Z_{1},N_{1},E_{1},\beta$) to ($Z_{1},N^{'}_{1},E^{'}_{1},\beta$)], and $d_{Z_{1},Z_{1}}$ denotes the microscopic dimension corresponding to the macroscopic state ($Z_{1},N_{1},E_{1}$). The sum is taken over all possible proton and neutron numbers that fragment ($Z^{'}_{1}$,$N^{'}_{1}$) may take, but only one nucleon transfer is considered in the model with the relations $Z^{'}_{1}=Z_{1}\pm1$ and $N^{'}_{1}=N_{1}\pm1$. 

The excited DNS opens a valence space in which the valence nucleons have a symmetrical distribution around the Fermi surface. Only the particles at the states within the valence space are actively for nucleon transfer. 
The transition probability is related to the local excitation energy and nucleon transfer, which is microscopically derived from the interaction potential in valence space as described as \cite{Chen_2017,1975ZPhyA.274.241N}.
\begin{eqnarray}
\label{trw}
&&W_{Z_{1},N_{1},\beta;Z_{1}^{\prime},N_{1},\beta ^{\prime}}= \frac{\tau_{\rm mem}(Z_{1},N_{1},\beta,E_{1};Z_{1}^{\prime},N_{1} \beta ^{\prime},E_{1}^{\prime})}{d_{Z_{1},N_{1}} d_{Z_{1}^{\prime},N_{1}}\hbar^{2}}  \times   \nonumber \\
&&\sum_{ii^{\prime}}|\langle  Z_{1}^{\prime},N_{1},E_{1}^{\prime},i^{\prime}|V|Z_{1},N_{1},E_{1},i \rangle|^{2}.
\end{eqnarray}

A similar method is used to calculate the neutron transition coefficient.
The relaxation time is calculated using the method of deflection function\cite{PhysRevC.27.590}, typically several hundred $10^{-22}$ s.
Memory time $\tau_{\rm mem}$ and $V$ interaction elements can be seen in the Ref\cite{Chen_2017}.

The motion of nucleons in the interacting potential is governed by the single-particle Hamiltonian\cite{PhysRevC.76.044606,Li_2003} as 
\begin{eqnarray}
H(t) = H_0(t) + V(t)
\end{eqnarray}
with
\begin{eqnarray}
H_0(t) && = \sum _K\sum_{\nu_K} \varepsilon_{\nu_K}(t)\alpha^+_{\nu_K}(t)\alpha_{\nu_K}(t)  \\
V(t) && = \sum_{K,K^{'}} \sum_{\alpha_K,\beta_{K'}} u_{\alpha_K,\beta_{K'}}\alpha^+_{\alpha_K}(t)\alpha_{\beta_K}(t)  \\ && = \sum_{K,K'}V_{K,K'}(t) \nonumber
\end{eqnarray}
Here the indices $K$, $K'$ ($K,K'$ = 1, 2) denote the fragment 1 and 2. The quantities $\varepsilon_{\rm \nu K}$ and $u_{\rm \alpha_K,\beta_{\rm K'}}$ represent the single particle energies and the interaction matrix elements, respectively. The single particle states are defined with respect to the centers of the interacting nuclei and are assumed to be orthogonalized in the overlap region. So the annihilation and creation operators are dependent on time. The single particle matrix elements are parameterized by
\begin{eqnarray}
&& u_{\rm \alpha_K,\beta_K'} =  \\ &&U_{\rm K,K'}(t) \left\{ \exp \left[- \frac{1}{2}( \frac{\varepsilon_{\alpha_{\rm K} }(t)  - \varepsilon_{\beta_{\rm K} }(t)}{\Delta_{\rm K,K'}(t)})^2 \right]  - \delta_{\rm \alpha_K,\beta_{K'}} \right\}  \nonumber
\end{eqnarray}
The detailed calculation of these parameters and the mean transition probabilities were described in Ref. \cite{PhysRevC.76.044606,Li_2003}.
\begin{eqnarray}
\label{vas}
\Delta \varepsilon_{\rm K} = \sqrt{\frac{4\varepsilon^*_{\rm K} }{g_{\rm K} }},\quad
\varepsilon^*_{\rm K} =\varepsilon^*\frac{A_{\rm K}}{A}, \quad
g_{\rm K} = A_{\rm K} /12,
\end{eqnarray}
Where the $\varepsilon^*$ is the local excitation energy of the DNS. The microscopic dimension for the fragment ($Z_{K},N_{K}$) is evaluated by the valence states $N_K$ = $g_K\Delta\varepsilon_K$ and the valence nucleons $m_K$ = $N_K/2$ ($K=1,2$) as
\begin{eqnarray}
\label{dmn}
 d(m_1, m_2) = {N_1 \choose m_1} {N_2 \choose m_2}.
\end{eqnarray}
The local excitation energy $E_{1}$ was derived by the dissipation energy coulpled to potential energy surface (PES) of the relative motion of DNS. The excitation energy in the equilibrium stage is owned by the fragments divided by mass. The angular momentum of the main fragment is determined by the moment of inertia. The local excitation energy evaluated by\cite{Chen_2017,1975ZPhyA.274.241N}
\begin{eqnarray}
\label{lee}
\varepsilon^{\ast}(t)=E^{\rm diss}(t)-\left(U(\{\alpha\})-U(\{\alpha_{\rm EN}\})\right).
\end{eqnarray}
The entrance channel quantities $\{\alpha_{\rm EN}\}$ include the proton and neutron numbers, quadrupole deformation parameters and orientation angles being $Z_{ \rm P}$, $N_{\rm P}$, $Z_{\rm T}$, $N_{\rm T}$, $R$, $\beta_{\rm P}$, $\beta_{\rm T}$, $\theta_{\rm P}$, $\theta_{\rm T}$ for projectile and target nuclei with the symbols of $P$ and $T$, respectively. The interaction time $\tau_{\rm int}$ is obtained from the deflection function method \cite{Wolschin1978AnalysisOR}. 
The energy dissipated into the DNS increase exponentially.
The potential energy surface (PES) of the DNS is evaluated by
\begin{eqnarray}\label{dri}
U_{\rm dr}(t) = Q_{\rm gg}+V_{\rm C}(Z_1,N_1;\beta_1,Z_2,N_2,\beta_2,t)     \nonumber   \\ 
+ V_{\rm N}(Z_1,N_1,\beta_1;Z_2,N_2,\beta_2,t) + V_{\rm def}(t)
\end{eqnarray}
with
\begin{eqnarray}\label{vcn}
V_{\rm def}(t) = \frac{1}{2} C_1 (\beta_1 - \beta' _T (t) )^2 + \frac{1}{2} C_2 (\beta_2 - \beta' _P (t) )^2      \\
C_i = (\lambda-1) { (\lambda+2) R^2_{\rm N} \delta - \frac{3}{2\pi}} \frac{Z^2e^2}{R_{\rm N}(2\lambda+1)}.
\end{eqnarray}
Where, the $Q_{\rm gg}$ derived by the negative binding energies of the fragments $(Z_{\rm i},N_{\rm i})$ were calculated by liquid drop model plus shell correction\cite{MOLLER20161}. The $\theta_{i}$ denotes the angles between the collision orientations and the symmetry axes of the deformed nuclei. $V_{\rm C}$ and $V_{\rm N}$ were calculated by the Wong formular\cite{PhysRevLett.31.766} and double-folding potential\cite{PhysRevC.69.024610}, respectively.
$V_{\rm def}(t)$ is the deformation energy of DNS at the reaction time $t$.
The evolutions of quadrupole deformations of projectile-like and target-like fragments undergo from the initial configuration as
\begin{eqnarray}\label{qde}
\beta' _{\rm T} (t) = \beta _ {\rm T} \exp(-t/\tau_{\rm \beta}) + \beta_1 [ 1 - \exp(-t/\tau_{\rm \beta})],      \nonumber \\
\beta' _{\rm P} (t) = \beta _ {\rm P} \exp{(-t/\tau_{\rm \beta})} + \beta_2 [ 1 - \exp(-t/\tau_{\rm \beta})]
\end{eqnarray}
with the deformation relaxation is $\tau_{\rm \beta}=4\times10^{-21} \ s$.

The total kinetic energy (TKE-mass) of the primary fragment is evaluated by the following expression.
\begin{equation}\label{tke}
 \rm{TKE}  = E_{\rm c.m.} + Q_{ \rm gg} - E^{\rm diss},
\end{equation}
where $Q_{\rm gg} = M_{\rm P} + M_{\rm T} - M_{\rm PLF} -M_{\rm TLF}$ and $E_{\rm c.m.}$ is the incident energy in the center of mass frame. The mass $M_{\rm P}$, $M_{\rm T}$, $M_{\rm PLF}$ and $M_{\rm TLF}$ correspond to projectile, target, projectile-like fragment and target-like fragment, respectively. 

The survival probability $W_{\rm sur}(E_{1},J_{1},s)$ is particularly important in evaluation of the cross section, which is usually calculated with the statistical model. The physical process in understanding the excited nucleus is clear. However, the magnitude of survival probability were strongly dependent on the ingredients in the statistical model, such as level density, separation energy, shell correction, fission barrier etc. The excited fragments were cooled by evaporating $\gamma$-rays, light particles (neutrons, protons, $\alpha$ etc) in competition with fission\cite{Chen_2016}.
the probability in the channel of evaporating the $x-$th neutron, the $y-$th proton and the $z-$ alpha is expressed as
\begin{eqnarray}
&&W_{\rm sur}(E^*_{\rm 1},x,y,z,J) =  P(E^*_{1},x,y,z,J)             \nonumber   \\
&& \times   \prod^x_{i=1} \frac{\Gamma_n(E^*_{\rm i},J)}{\Gamma_{\rm tot}(E^*_{\rm i},J)}
\prod^y_{j=1} \frac{\Gamma_p(E^*_{\rm j},J)}{\Gamma_{\rm tot}(E^*_{\rm j},J)}
 \prod^z_{k=1} \frac{\Gamma_{\alpha}(E^*_{\rm k},J)}{\Gamma_{\rm tot}(E^*_{\rm k},J)}.
\end{eqnarray}
Here the $E^*_{1}$, $J$ are the excitation energy evaluated from the mass table in Ref. \cite{MOLLER20161} and the spin of the excited nucleus, respectively. The total width $\Gamma_{\rm tot}$  is the sum of partial widths of particle evaporation, $\gamma$-emission and fission. The excitation energy $E^*_s$ before evaporating the $s$-th particle is evaluated by
\begin{equation}
E^*_{s+1} = E^*_{\rm s} - B^n_{\rm i} - B{\rm ^p_j} - B^{\alpha}_{\rm k} - 2T_{\rm s}
\end{equation}
with the initial condition $E^*_1$ and $\rm s=i+j+k$. The $B^n_i$, $B^p_j$, $B^{\alpha}_k$ are the separation energy of the $i$-th neutron, $j$-th proton, $k$-th alpha, respectively. The nuclear temperature $T_i$ is given by $E^*_{\rm i} = aT{\rm _i^2}-T_{\rm i}$ with $a$ being the level density parameter.
The fission width and particle decay width were calculated by Weisskopf evaporation theory and Bohr-Wheeler formula, respectively. The realization probability $P(E^*_{1},x,y,z,J)$ was calculated by the Jackson formula\cite{doi:10.1139/p56-087}.
\section{Results and discussion}\label{sec3}
\begin{figure}[htb]
\includegraphics[width=1.\linewidth]{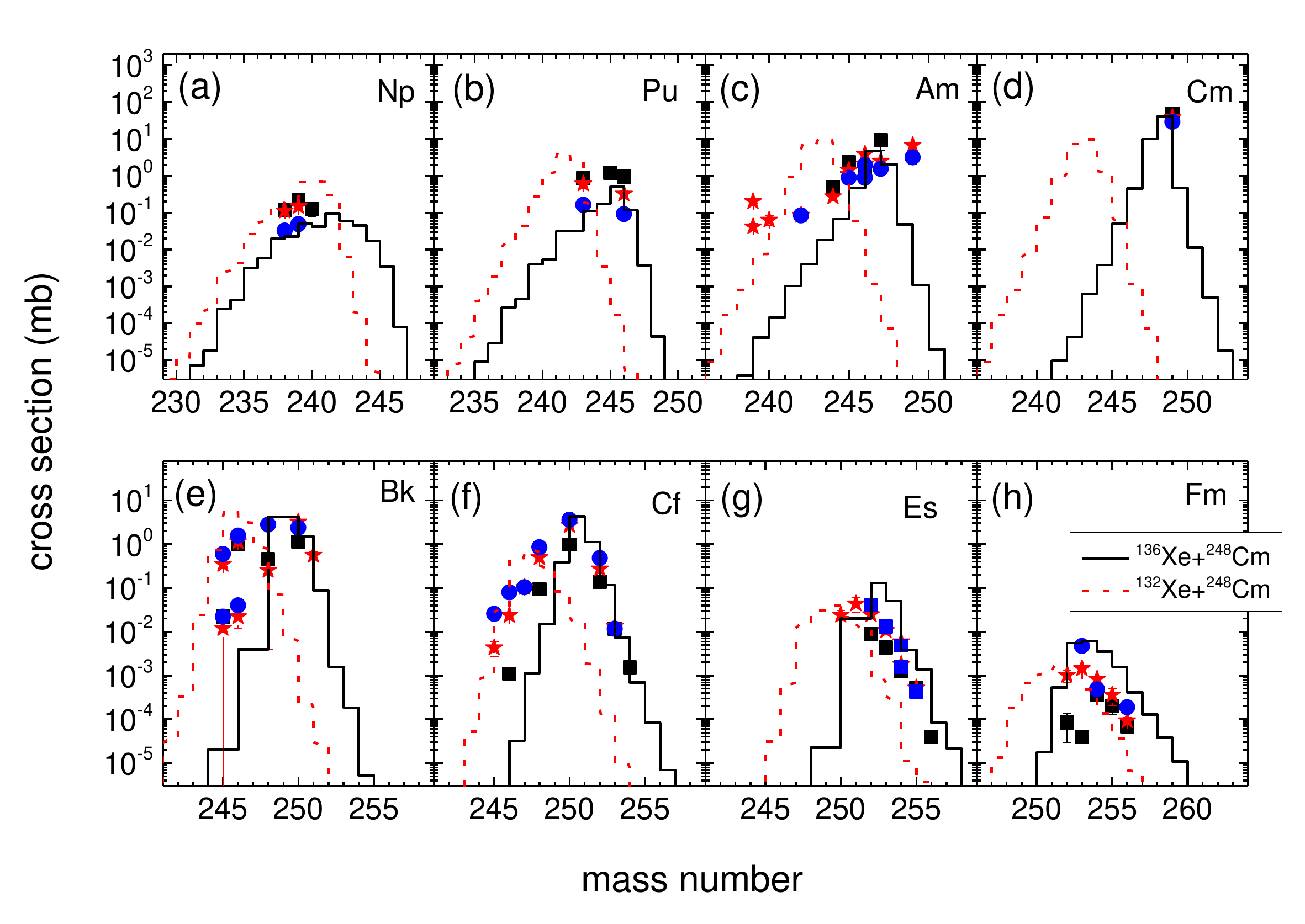}
\caption{\label{fig1} (Color online) Calculation and experiment results of production cross-secions of actinide isotopic chains with Z = 93-100 in reactions of $^{129,132,136}$Xe + $^{248}$Cm at $E_{\rm lab}$ = 699-790 MeV.
The available exprimental data are taken from \cite{PhysRevC.35.204,PhysRevC.33.1315}, marked by solid black square for $^{136}$Xe induced reactions, solid red circle for $^{132}$Xe induced reactions, solid blue star for $^{129}$Xe induced reactions. Our calculations for $^{136}$Xe induced reactions were shown by solid black lines, $^{132}$Xe induced reactions shown by dash red lines.}
\end{figure} 
We calculated the production cross-sections of actinide istopes chains with atomic number $Z = 93-100$ in collsions of $^{132,136}$Xe + $^{248}$Cm at incident energy $E_{\rm lab}$ = 699-790 MeV, as shown in Fig. \ref{fig1}. Comparing to the available experimental data of $^{129,132,136}$Xe + $^{248}$Cm which were represented by solid red star, solid blue circle and solid black squares with error bars, respectively, our calculation of $^{136}$Xe + $^{248}$Cm marked by solid black lines and $^{132}$Xe + $^{248}$Cm marked by solid red lines could basically reproduce the tendency of cross-sections distribution of actinide isotopic chains. From experimental data\cite{PhysRevC.35.204,PhysRevC.33.1315}, it was found that projectiles $^{129,132,136}$Xe isotopes induced reactions with target $^{248}$Cm gave actinide products which have a large overlap distribution area in the neutron-rich region. It was not distinguishable clearly as we expected. From our calculation in term of deep-inelastic mechanism, relative proton-rich projectiles $^{132}$Xe induced reactions tend to shift to proton-rich region, compared to experimental results. According to the data in Fig. \ref{fig1}, target-like fragments have production cross-sections of magnitude level from 100 millibarn to 10 nanobarn. To far off the target, formation cross section of target-below products decreas more slowly than transtarget products, which indicate quasi-fission was dominant in this collsions relatively.
\begin{figure}[htb]
\includegraphics[width=1.\linewidth]{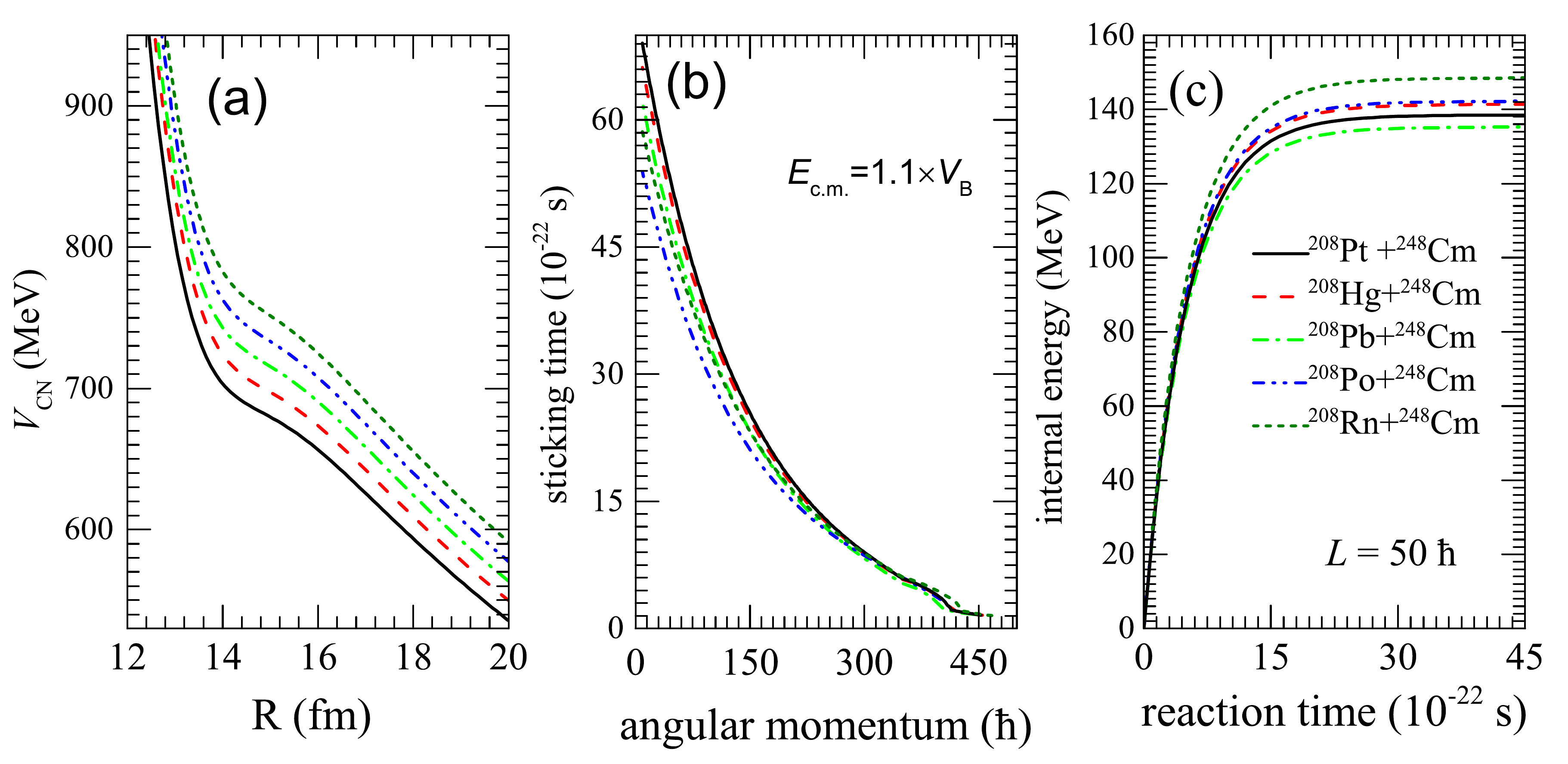}
\caption{\label{fig2} (Color online) The solid black, red, blue, green, olive lines indicate the interaction potential of the tip-tip collisions as a function of surface distance for reactions of projectiles $^{208}$Hg, $^{208}$Pb, $^{208}$Po, $^{208}$Pt and $^{208}$Rn induced reactions with target $^{248}$Cm, in panel (a);
In panel (b), it shows distributions of reaction time to angular momentum of collisions for these five reaction systems at incident energy $E_{\rm c.m.} = 1.1 \times V_{\rm B}$, which decrease exponentially with angular momentum increasing.
For given angular momentum $L =  50 \hbar$ in these five colliding systems, these internal excitation energies increase with reaction time exponentially in panel (c).}
\end{figure}

To investigate the competition of Coulomb repulsive potential and shell effect in MNT reactions, we calculated reactions of isobaric projectiles with $A=208$ bombarding on targets $^{248}$Cm and $^{232}$Th at incident energy $E_{\rm c.m.}=1.1\times V_{\rm B}$ within the framework of the dinuclear system model (DNS). Calulations detial for these collisions would be shown below.
Interaction potential between colliding partners were combined by Coulomb potential and nucleus-nucleus potential. In Fig. \ref{fig2} (a), interaction potential $V_{\rm CN}$ for $^{208}$Pt + $^{248}$Cm, $^{208}$Hg + $^{248}$Cm, $^{208}$Pb + $^{248}$Cm, $^{208}$Po + $^{248}$Cm and $^{208}$Rn + $^{248}$Cm reactions were marked by solid black, dash red, dash-dot blue, dash-dot-dot green, short dash olive lines, respectively. The tendency of $V_{\rm CN}$ distributions for these collisions were similar. The larger Coulomb potential led to larger interaction potential $V_{\rm CN}$. $V_{\rm CN}$ increase exponentially with distance R decreasing, in the attraction region of nuclear force where it increase slowly. Nucleon transfer happended at the touch configuration. Based on deflection function, sticking time of colliding partners were calculated for all impact parameters\cite{PhysRevC.27.590}, shown in Fig. \ref{fig2} (b), which decrease expotentially with angular momentum increasing. 
In these cllisions, relative larger Coulomb potential cause the longer sticking time with the fixed impact parameter.
During the sticking time, kinetic energy dissipating into composite system to heat up with internal excitation energy, which increase with reaction time expotentially and reach equilibrium around 2$\times 10^{-21}$ s, shown in Fig. \ref{fig2} (c).

\begin{figure}[htb]
\includegraphics[width=1\linewidth]{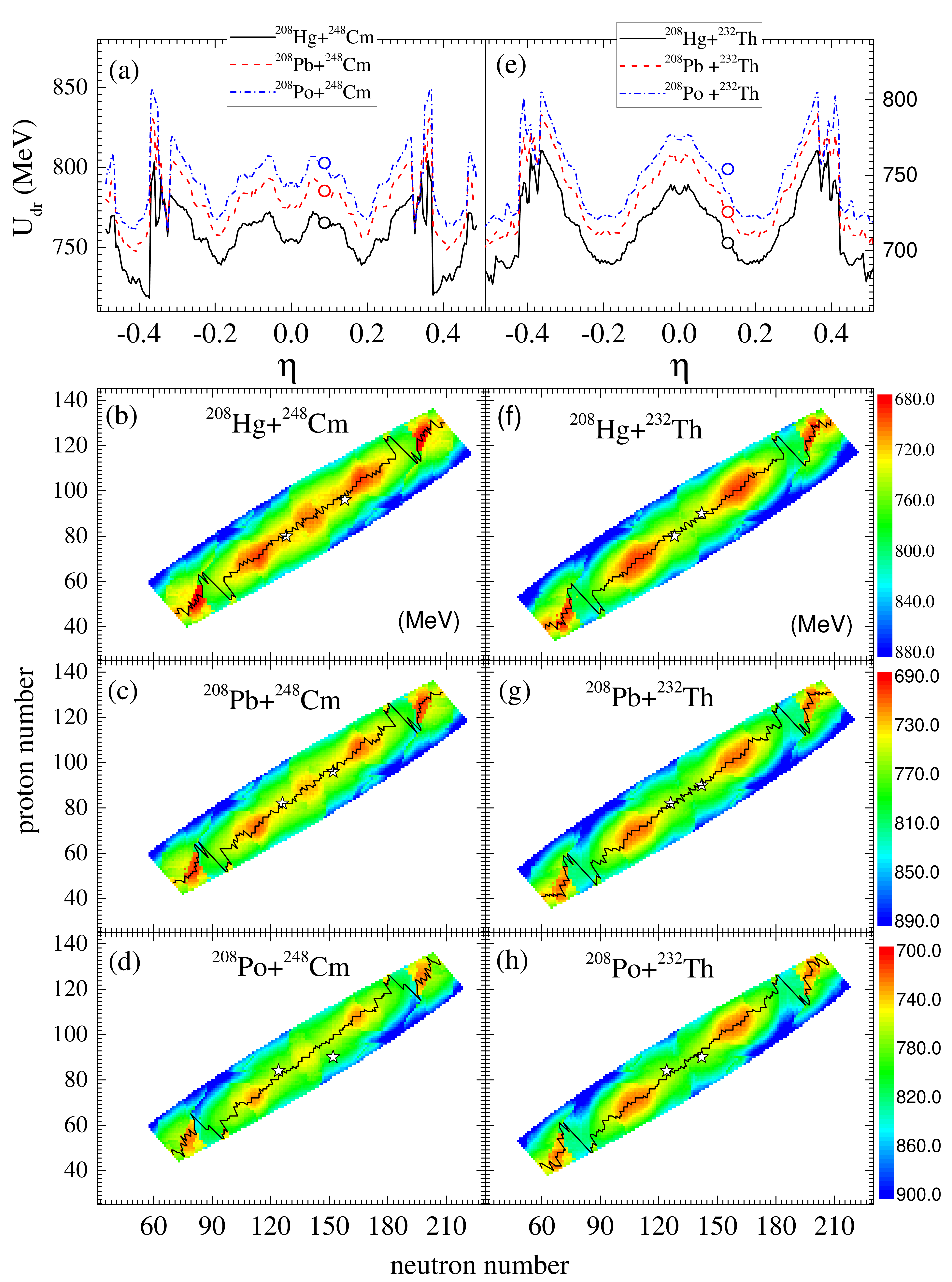}
\caption{\label{fig3} (Color online) Potential energy surface and driver potentials of projectiles $^{208}$Hg, $^{208}$Pb and $^{208}$Po induced reactions with targets $^{248}$Cm and $^{232}$Th at tip-tip collisions were listed in Fig. \ref{fig3}. 
$^{208}$Hg, $^{208}$Pb and $^{208}$Po induced reactions were represented by solid black, dash red and dash-dot blue lines in panels (a) and (e), respectively.
Potential energy surfaces for these collsions were shown in panels (b), (c), (d), (f), (g), (h), respectively. Open stars stand for projectile-target injection points.
These solid black lines were valley trajectories in two-dimensions potential energy surface.}

\end{figure}

After capture for these colliding partners, the dissipating kinentic energy coupled to angular momentum in DNS enable them to diffuse along potential energy surface (PES), followed by nucleons rearrangement between the colliding partners, which was calculated by sloving a set of master equations. PES and driving potential were calculated by Eq.(\ref{dri}) which were the composition of Coulomb potential, binding energy and nuclear force, computed by Wong formula, liquid-drop model plus shell correction, and double folding method\cite{FENG200650}, respectively.
Driving potential of projectiles $^{208}$Hg, $^{208}$Pb and $^{208}$Po on targets $^{248}$Cm and $^{232}$Th at tip-tip collision with fixed distance plotted as function of mass asymmetry $\eta$ respected to $\eta = (A_{\rm T}-A_{\rm P})/(A_{\rm T}+A_{\rm P})$, illustrated in Fig.\ref{fig3}(a)(e), represented by solid black, dash red, dash-dot blue lines, respectively. 
Arrow lines and open stars stand for projectile-target injection points. 
From panels (a) and (e), it was found that the tendency of driving potential trajectories for these collisions were similar. There are two pockets appear at $\eta = 0.2, 0$ for deriving potential of target $^{248}$Cm-based reactions. One pocket in deriving potentials for target $^{232}$Th-based reactions appears at $\eta = 0.2$. Neutron subshell number $N=162$ might play a crucial role in pockets formation.
According to PESs of these reactions, targets prefer to pick up nucleons from projectiles, compared to lose nucleons to projectiles. Namely, deep-inelastic reaction mechanism was dominant, instead of quasifission. 
For the projectiles $^{208}$Po far from $\beta -$ stable line, their injection points were far off their deriving potential trajectories. When diffusion begins, it tend to driving potential trajectorie rapidly.
In general, based on PES, the spectra distribution trend of each isotope chain have been predicted roughly.

\begin{figure}[htb]
\includegraphics[width=1\linewidth]{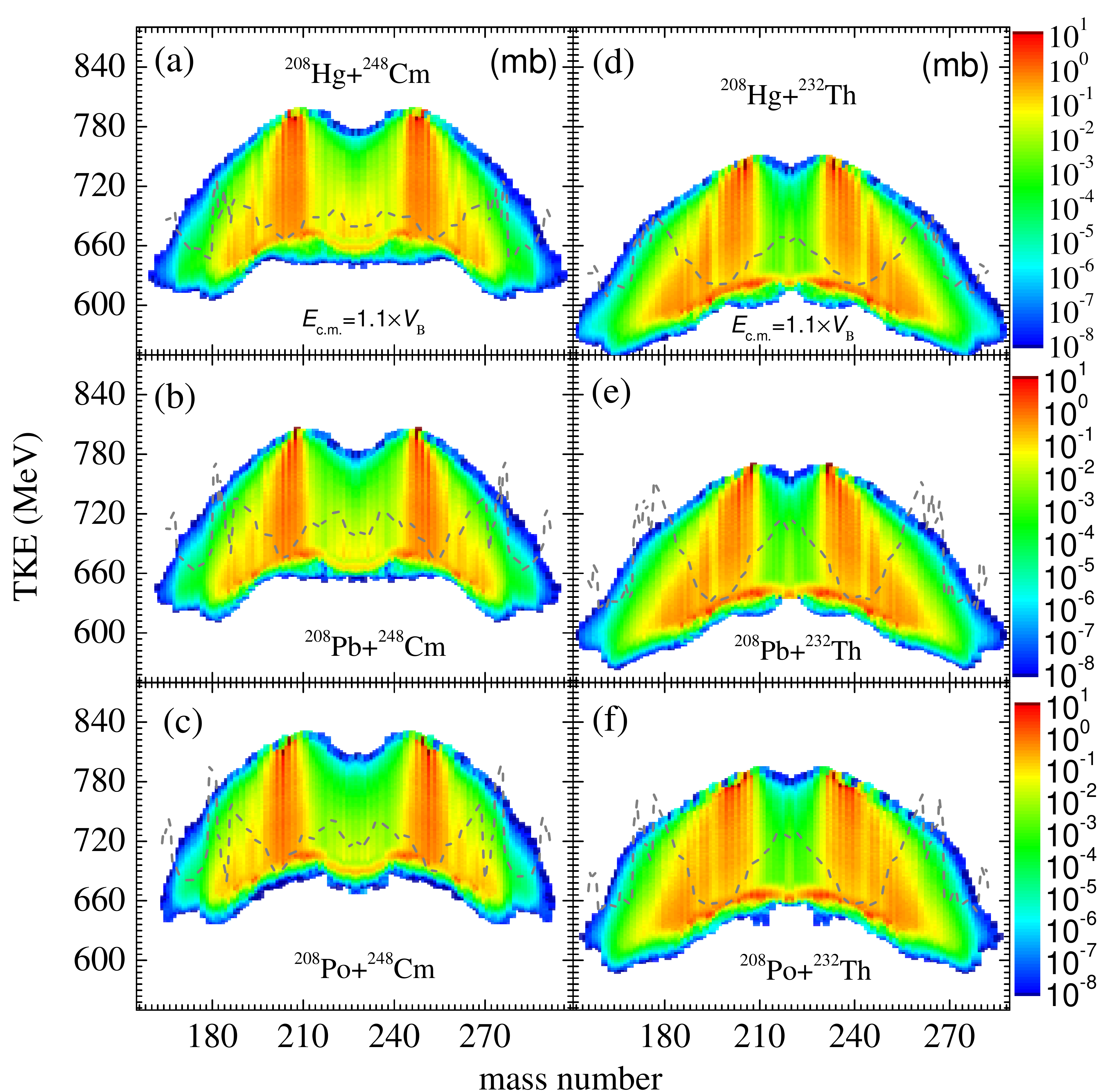}
\caption{\label{fig4}(Color online) Calculated TKE-mass distribution of primary reaction products in head-on collisions of projectiles $^{208}$Hg, $^{208}$Pb and $^{208}$Po induced reactions with targets $^{248}$Cm and $^{232}$Th at $E_{\rm c.m.} = 1.1 \times V_{\rm B}$ were shown in panels (a), (b), (c), (d), (e) and (f), respectively, where driving potential trajectories were added in. }
\end{figure}

Production probabilities of primary fragments with excitation energies were derived by solving a set of master equations, which were classified by mass number and kinetic energy, derived by Eq. (\ref{tke}), illustrated in Fig. \ref{fig4}, where driving potential trajectaries were added as solid grey lines. 
From Fig. \ref{fig4}, it was found that two peaks in large kinetic region located around projectile-target injection points and 
cross-sections prefer to populate around pockets of driving potential trajectories. For all the reactions of projectiles $^{208}$Hg, $^{208}$Pb and $^{208}$Po induced with targets $^{248}$Cm and $^{232}$Th at incident energy $E_{\rm c.m.} = 1.1 \times V_{\rm B}$ gave the similar shape of TKE-mass distributions which have symmetric and broad distributions. 
The TKE-mass distribution is very wide in the kinetic range of 500 to 800 MeV and the mass region of 160 to 280, which might be expect to transfer more than 30 nucleons.

\begin{figure}[htb]
\includegraphics[width=1\linewidth]{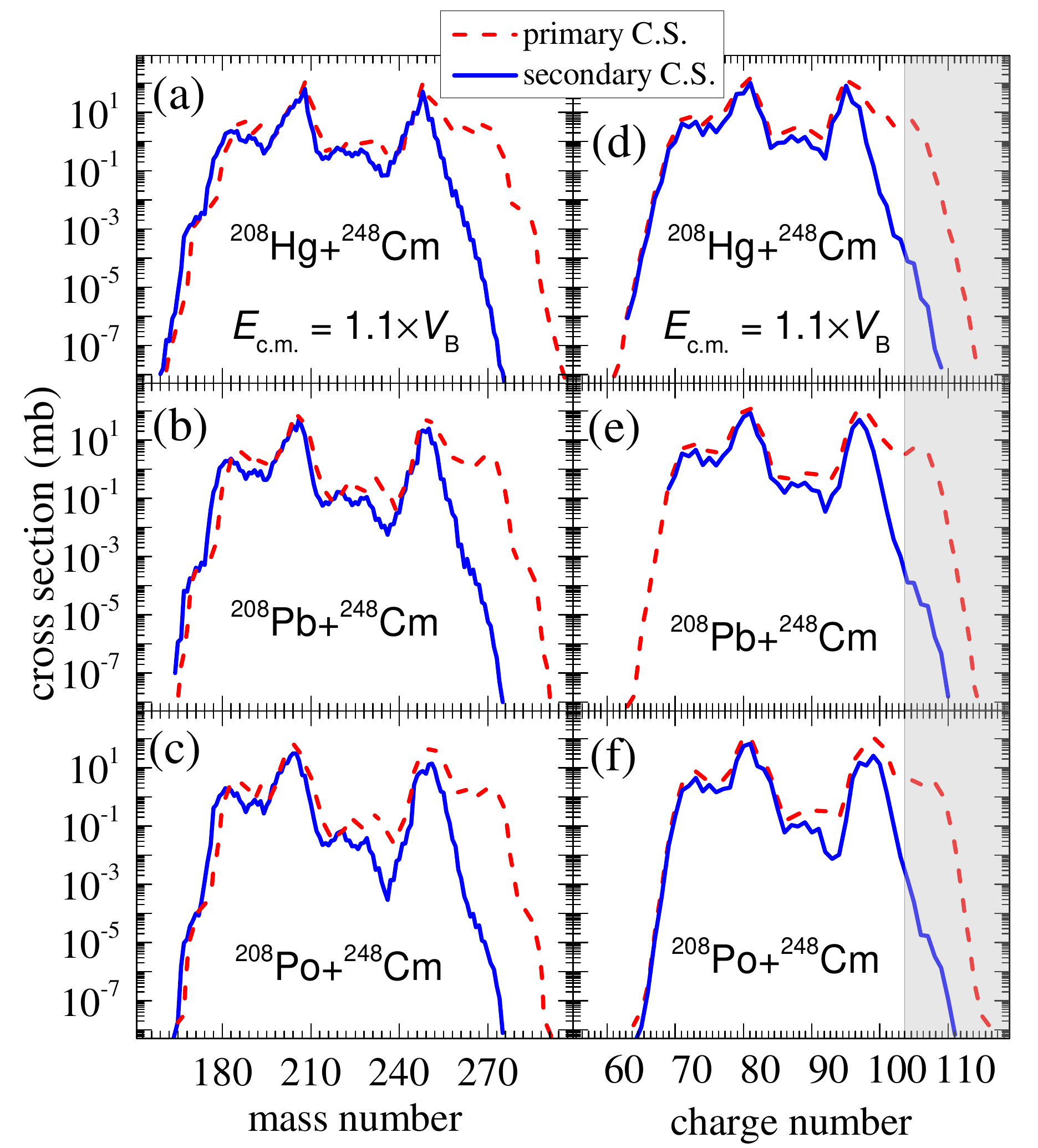}
\caption{\label{fig5}(Color online)
The calculated primary and secondary yields mass and charge distribution for $^{208}$Hg, $^{208}$Pb and $^{208}$Po induced reactions with targets $^{248}$Cm at $E_{\rm c.m.} = 1.1 \times V_{\rm B}$ were listed in panels (a), (b), (c), (d), (e) and (f), respectively. 
Dash red and solid blue lines represented primary and secondary yields. The superheavy region ($Z > $104) were shown by rectangular shadow. }
\end{figure}

Based on the statistics evaporation program, survival probability of excited primary fragments has been calculated, gave the production cross-sections of secondary fragments. 
Production yields of primary and secondary fragments as functions of mass number and charge number in collisions of projectiles $^{208}$Hg, $^{208}$Pb and $^{208}$Po induced reactions with targets $^{248}$Cm at $E_{\rm c.m.} = 1.1 \times V_{\rm B}$ were listed in Fig. \ref{fig5} (a), (b), (c), (d), (e) and (f), respectively. Solid blue and dash red lines indicate secondary fragments and primary fragments. Superheavy nuclei region were covered by rectangular shadows.
From Fig. \ref{fig5}, it was found that primary fragments could cover very large charge region, even access the superheavy region and secondary fragments were depressed by de-excitation strongly. Because of highly excited primary transtarget fragments with small fission barrier led to fisse easily. 
Predicted cross-sections of superheavy nuclei with $Z = 104-116$ were over 10 picobarn, where neutron subshell $N=162$ might play a crucial role in, especially in collision of $^{208}$Po + $^{248}$Cm. 

\begin{figure}[htb]
\includegraphics[width=1\linewidth]{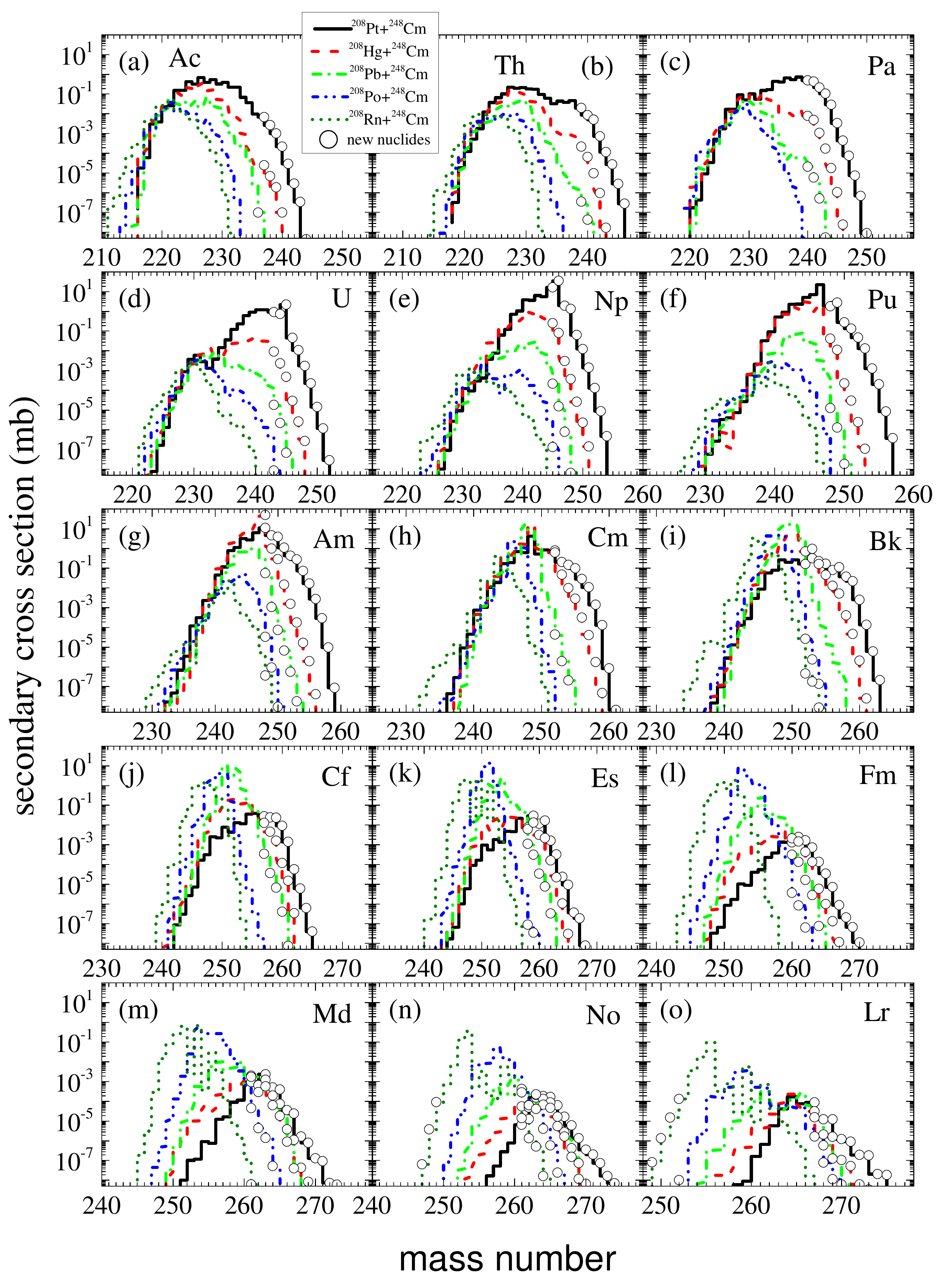}
\caption{\label{fig6}(Color online)
Predicted cross-sections of isotopic distribution of target-like fragments with $Z = 89-103$ in collisions of projectiles Pt, Hg, Pb, Po, Rn with mass number $A = 208$ bombarding on target $^{248}$Cm at $E_{\rm c.m.} = 1.1 \times V_{\rm B}$, correspond to solid black, dash red, dash-dot green, dash-dot-dot blue, and short dash olive lines, respectively, where predicted new actinide isotopes marked by open circles were added in. }
\end{figure}

Calculated secondary production cross-sections of actinide target-like fragments of Actinium, Thorium, Protactinium, Uranium, Neptunium, Plutonium, Americium, Curium, Berkelium, Californium, Einsteinium, Fermium, Mendelevium, Nobelium, Lawrencium isotopes in collisions of projectiles $^{208}$Pt, $^{208}$Hg, $^{208}$Pb, $^{208}$Po, $^{208}$Pt, $^{208}$Rn bombarding on targets $^{248}$Cm at $E_{\rm c.m.} = 1.1 \times V_{\rm B}$, correspond to solid black, dash red, dash-dot green, dash-dot-dot blue and short dash olive lines, respectively, illustrated in Fig. \ref{fig6}. 
From Fig. \ref{fig6}, it was found that collisions with the smaller Coulomb force prefer to shift to neutron-rich region and with the larger Coulomb force tend to neutron-deficient area. 
Massive unknown actinide isotopes have been predicted by all of the reactions $^{208}$Pt+$^{248}$Cm, $^{208}$Hg+$^{248}$Cm, $^{208}$Pb+$^{248}$Cm, $^{208}$Po+$^{248}$Cm, $^{208}$Pt+$^{248}$Cm, $^{208}$Rn+$^{248}$Cm. For new neutron-rich actinide isotopes, $^{208}$Pt+$^{248}$Cm reactions prefer to produce the largest cross-sections. However, $^{208}$Pt is still a unknown nuclide. 
The $^{208}$Pb induced reactions could be used to synthesize actinide isotopes, such as, 
for actinium, producing unknown-yet $^{237,238,239,240}$Ac as 24 nb, 7 nb, 7 nb, 0.1 nb, and 
for thorium, producing unknown-yet $^{239,240,241,242}$Th as 216 nb, 64 nb, 9 nb, 0.9 nb, and 
for protactinium, producing unknown-yet $^{240,241,242,243,244,245}$Pa as 7790 nb, 4890 nb, 1080 nb, 171 nb, 17 nb, 1 nb, and  
for uranium, producing unknown-yet $^{243,244,245,246}$U as 9190 nb, 2290 nb, 478 nb, 28 nb, and 
for neptunium, producing unknown-yet $^{245,246,247,248,249,250}$Np as 0.25 mb, 0.08 mb, 3070 nb, 371 nb, 26 nb, 2 nb, and
for plutonium, producing unknown-yet $^{248,249,250,251,252,253}$Pu as 0.04 mb, 0.01 mb, 226 nb, 30 nb, 2.4 nb, 0.1 nb, and
for americium, producing unknown-yet $^{248,249,250,251,252,253,254,255}$Am as 49 mb, 1 mb, 0.3 mb, 0.01 mb, 1530 nb, 176 nb, 16 nb, 2 nb and
for curium, producing unknown-yet $^{252,253,254,255,256,257,258}$Cm as 0.6 mb, 0.04 mb, 0.01 mb, 2820 nb, 314 nb, 40 nb, 1.7 nb, and
for berkelium, producing unknown-yet $^{252,253,254,255,256,257,258,259,260}$Bk as 0.7 mb, 0.9 mb, 0.16 mb, 0.05 mb, 0.01 mb, 5200 nb, 413 nb, 46 nb, 2 nb, and
for californium, producing unknown-yet $^{257,258,259,260,261,262}$Cf as 0.01 mb, 4580 nb, 1580 nb, 118 nb, 15 nb, 0.3 nb, and 
for einsteinium, producing unknown-yet $^{258,259,260,261,262,263,264}$Es as 6200 nb, 0.01 mb, 3110 nb, 2190 nb, 162 nb, 22 nb, 1 nb, and 
for fermium, producing unknown-yet $^{260,261,262,263,264,265}$Fm as 2100 nb, 1450 nb, 439 nb, 135 nb, 6 nb, 1 nb.
$^{208}$Pb+$^{248}$Cm reactions prefer to synthesize new actinides with $Z=101, 102, 103$, 
for mendelevium, producing unknown-yet $^{261,262,263,264,265,266,267}$Md as 1940 nb, 625 nb, 1250 nb, 194 nb, 73 nb, 4 nb, 1 nb and 
for nobelium, producing unknown-yet $^{261,262,263,264,265,266,267,268}$No as 54 nb, 64 nb, 205 nb, 131 nb, 86 nb, 16 nb, 2.6 nb, 0.4 nb and 
for lawrencium, producing unknown-yet $^{267,268,269}$Lr as 15 nb, 0.5 nb, 0.1 nb.
It should be noticed that unknown actinide products are highly depend on Coulomb force.
$^{208}$Rn+$^{248}$Cm reactions prefer to produce new neutron-deficient actinide isotopes with the largest cross-sections.
The open circles represent new neutron-rich actinide nuclides and new proton-rich actinide nuclides.

\begin{figure}[htb]
\includegraphics[width=1.\linewidth]{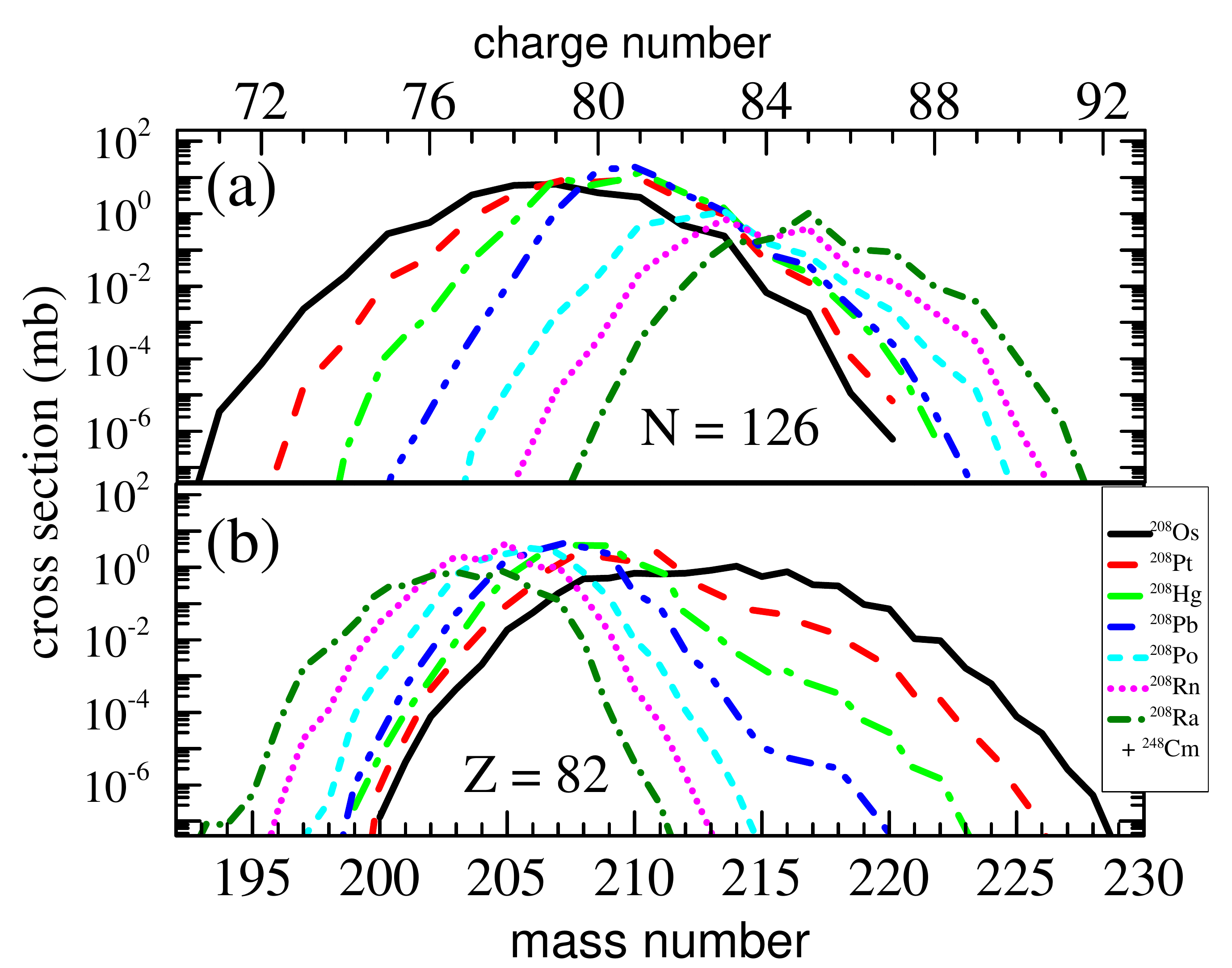}
\caption{\label{fig7}(Color online)
For reactions of $^{208}$Pt, $^{208}$Hg, $^{208}$Pb, $^{208}$Po, $^{208}$Pt, $^{208}$Rn and $^{208}$Ra colliding on $^{248}$Cm, production cross-sections of fragments with magic number $N = 126$ and $Z = 82$ were indicated by solid black, dash red, dash-dot green, dash-dot-dot blue, short dash cyan, dot magenta, short dash-dot olive lines, respectively.}
\end{figure}

\begin{figure*}[htb]
\includegraphics[width=.88\linewidth]{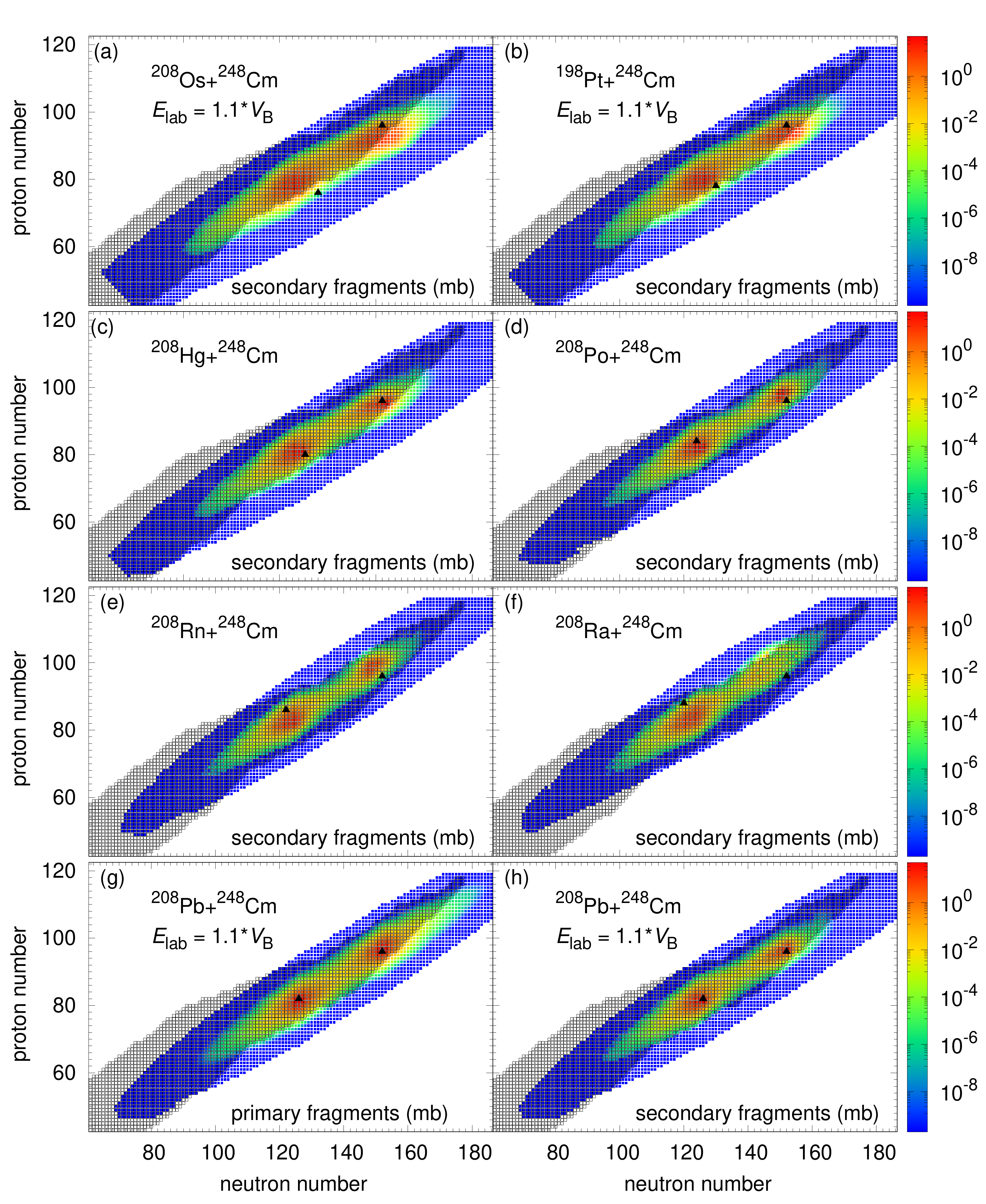}
\caption{\label{fig8}(Color online) 
The production cross section of secondary all the formed fragments in collisions of $^{208}$Pt + $^{248}$Cm, $^{208}$Hg + $^{248}$Cm, $^{208}$Pb + $^{248}$Cm, $^{208}$Po + $^{248}$Cm $^{208}$Rn + $^{248}$Cm and $^{208}$Ra + $^{248}$Cm at the incident energy $E_{\rm c.m.} = 1.1 \times V_{\rm B}$ and primary fragments of $^{208}$Pb + $^{248}$Cm were listed in $N-Z$ panels. 
Open stars stand for projectile-target injection points.}
\end{figure*}

All fragments with neutron shell number $N=126$ and proton shell number $Z=82$ were from these five colliding systems, shown in Fig. \ref{fig8} (a) and (b), respectively. The solid black, dash red, dash-dot green, dash-dot-dot blue, short dash cyan, dot magenta, short dash-dot olive lines indicate reactions $^{208}$Pt+$^{248}$Cm, $^{208}$Hg+$^{248}$Cm, $^{208}$Pb+$^{248}$Cm, $^{208}$Po+$^{248}$Cm, $^{208}$Pt+$^{248}$Cm and $^{208}$Rn+$^{248}$Cm, $^{208}$Ra+$^{248}$Cm orderly.
From Fig. \ref{fig7} (a), it was found that cross section of isotones with $N=126$ had a wider distribution in the more neutron-rich projectiles induced reactions, because, transfer neutrons happened in neutron-rich projectiles easily and it tend to shift to neutron-rich side. The relative proton-rich projectiles induced reactions gave the narrow cross section distribution, where protons were not transferred easily because of Coulomb barrier. The largest peaks for all these reactions were around proton shell number $Z = 82$, which show the proton shell effect on the production cross section of fragments formation.
From Fig. \ref{fig7} (b), production cross section of isotopes with $Z=82$ show the largest peaks around the neutron shell number $N=126$ which display the neutron shell effect. 

Figure \ref{fig8} show secondary production cross-sections of all the formed fragments in collisions of $^{208}$Os+$^{248}$Cm, $^{208}$Pt+$^{248}$Cm, $^{208}$Hg+$^{248}$Cm, $^{208}$Pb+$^{248}$Cm, $^{208}$Po+$^{248}$Cm, $^{208}$Rn+$^{248}$Cm, $^{208}$Ra+$^{248}$Cm and primary production cross-sections of $^{208}$Pb+$^{248}$C at the incident energy $E_{\rm c.m.} = 1.1 \times V_{\rm B}$ as $N-Z$ panel. 
From panels (g) and (h), it clearly show the de-excitation effect.
From panels (a), (b), (c), (d), (e), (f) and (h), it was found that massive new isotopes were predicted inculding neutron-rich and -deficient isotopes, even the superheavy nuclei. 
The projectile-target injection points and all the existed isotopes in the nuclide chart were represented by solid black up triangles, and open squares, respectively.

\section{Conclusions}\label{sec4}
Within the framework of dinuclear system model, production cross-sections of MNT fragments in reactions of projectiles of $^{208}$Os, $^{208}$Pt, $^{208}$Hg, $^{208}$Pb,$^{208}$Po, $^{208}$Rn, $^{208}$Ra,$^{132,136}$Xe bombarding on targets of $^{232}$Th and $^{248}$Cm around Coulomb barrier energies have been calculated systematically. To investigate the isospin diffusion on formation of actinide products in multinucleon transfer process, same mass number of projectiles with $A=208$ were selected. 
Our calculation of $^{132,136}$Xe + $^{248}$Cm have a nicely consistent with the available experimental data.
Sticking time of the colliding systems derived by deflection functions are highly dependent on the Coulomb force, especially at the small impact parameters. 
PES and TKE of these reactions are disscused, which could contribute to predict the tendency of cross-section diffusion. The relative large cross section from TKE appear at around the pockets in PES, where the neutron subshell $N=162$ exhibit evidently. 
The de-excitation process strongly depress the primary cross section of actinide isotopes up to four magnitude levels.
The production cross section of new actinides are highly dependent on the $N/Z$ ratio of the isobaric projectile.
It is found that Coulomb force coupled to shell effect play a crucial role in production cross-sections of actinides products in MNT reactions. 
The shell effect is shown on the production cross section of the isotones with $N=126$ and isotopes with $Z=82$.
Massive unknown heavy isotopes have been predicted with available cross-sections value by these five colliding systems, even for the superheavy nuclei with charge number $Z=104-110$.

\section{Acknowledgements}
This work was supported by National Science Foundation of China (NSFC) (Grants No. 12105241,12175072), NSF of Jiangsu Province (Grants No. BK20210788), Jiangsu Provincial Double-Innovation Doctor Program(Grants No. JSSCBS20211013) and University Science Research Project of Jiangsu Province (Grants No. 21KJB140026). This project was funded by the Key Laboratory of High Precision Nuclear Spectroscopy, Institute of Modern Physics, Chinese Academy of Sciences (CAS). This work was supported by the Strategic Priority Research Program of CAS (Grant No. XDB34010300).

%

\end{document}